\documentclass[conference]{IEEEtran}
\ifCLASSINFOpdf
\usepackage[pdftex]{graphicx}
\else
\fi
\usepackage[cmex10]{amsmath}
\usepackage{amssymb}
\usepackage{cite}

\IEEEoverridecommandlockouts \IEEEpubid{\makebox[\columnwidth]{ICNF2013\hfill 978-1-4799-0671-0/13/\$31.00~\copyright 2013~IEEE} \hspace{\columnsep}\makebox[\columnwidth]{ }}
\begin{document}
%
\title{Energy and particle number fluctuations in superconducting heterostructures}

\author{\IEEEauthorblockN{Ville J. Kauppila}
\IEEEauthorblockA{O.V. Lounasmaa Laboratory\\
Aalto University, 
Finland\\
Email: ville.kauppila@aalto.fi}
\and
\IEEEauthorblockN{Matti A. Laakso}
\IEEEauthorblockA{Institute for Theory of Statistical Physics,\\ RWTH Aachen University, 52056 Aachen, Germany\\
Email: laakso@physik.rwth-aachen.de}
\and
\IEEEauthorblockN{Tero. T. Heikkil\"a}
\IEEEauthorblockA{O.V. Lounasmaa Laboratory\\
Aalto University, Finland\\
Tero.Heikkila@aalto.fi}}


%


\maketitle

\begin{abstract}
We consider fluctuations of the energy on a mesoscopic island coupled to two leads. We use the Keldysh effective action formalism to derive the Langevin equation for the energy of the island in a very general setting and show how the Langevin equation for the case of uncorrelated tunneling events is derived from a more general one. As an application of the theory, we consider a superconducting island coupled to two normal metal leads and calculate the statistics of the temperature (in the quasiequilibrium case) and the number of quasiparticle excitations on the island.

\end{abstract}


%
\IEEEpeerreviewmaketitle

\section{Introduction}

Recently there has been a growing interest in thermodynamics \cite{RevModPhys.78.217} and especially fluctuations of thermodynamic quantities of nano-sized electronic systems \cite{PhysRevLett.108.067002,laakso2012theory,PhysRevLett.102.130605,kindermann2004statistics,averin2010violation,maisi2012single,PhysRevB.84.195117}. While in macroscopic systems, fluctuations of thermodynamic quantities are usually negligible, in nanoscale systems they can have a large effect on the transport properties of the system \cite{laakso2010giant}. Another source of interest to fluctuations is the possibility to use them to realize various types of Szilard engine or Maxwell's demon type \cite{maruyama2009colloquium,averin2011maxwell} machines in nanosystems.

In the field of electronic quantum transport, the system under consideration often consists of the actual device which we want to model and some leads which are coupled to the device and act as an environment. The energy and charge transport between the leads and the system is stochastic (by the nature of quantum mechanics) which causes fluctuations in the state of the system.

\begin{figure}[!ht]
\centering
\includegraphics[width=.99\columnwidth]{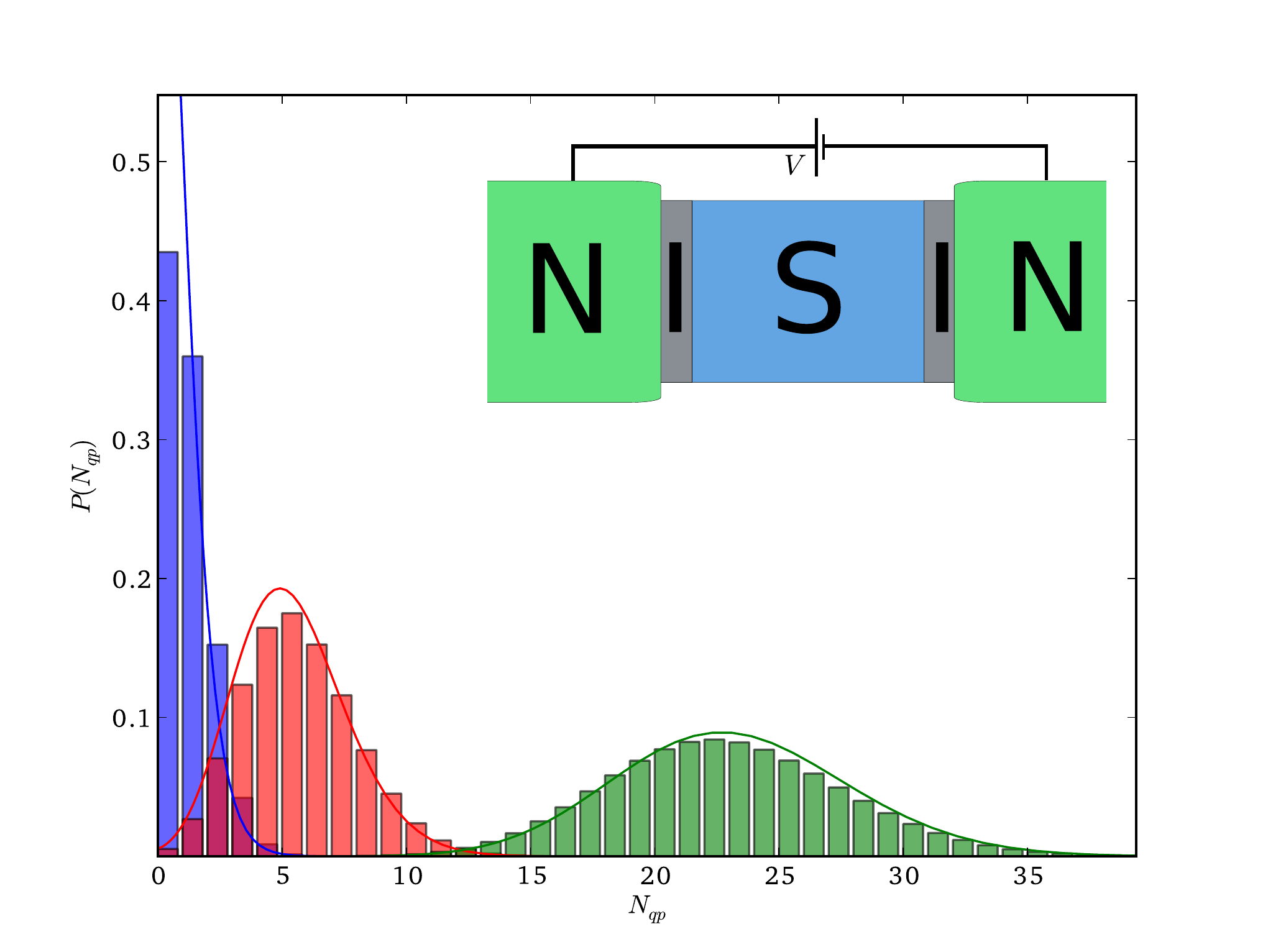}
\caption{\small Probability distribution for the quasiparticle number on a superconducting island connected to two normal metal leads assuming that each tunneling particle carries energy $\Delta$. The solid lines are the corresponding distributions calculated from the Fokker-Planck distribution \eqref{fokkerplanck} and substituting for the quasiparticle number an effective one from the BCS equilibrium relation between the quasiparticle number and the temperature. Blue, red and green curves are for $T_N/\Delta = 0.7, 0.8, 0.9$, respectively. The bias voltage is set to zero in all cases. Inset: Schematic picture of the setup.}
\label{fig:langevinNISIN}
\end{figure}

We present here a derivation for the Langevin equation for the energy of an island connected to two leads in a very general setting. As an example, we consider a superconducting island (S) connected to normal metal (N) leads (see the inset of Fig. \ref{fig:langevinNISIN}). In the case of our example, the fact that the tunneling rates have a strong peak at the edge of the superconducting gap leads to an approximate equivalence between the number of quasiparticle excitations and the energy on the island.

\section{Formalism}

In the full counting statistics formalism, the expectation values of observables and their moments (noise, etc.) are calculated from a generating function. The generating function is often written in the form (here and below, $\hbar = 1$)
\begin{equation}
\mathcal{Z} (\chi) = e^{S_{\rm{eff}}(\chi)} ,
\end{equation}
where $S_{\rm{eff}}$ is an effective action describing the system and $\chi$ is the counting field. The effective action can often be derived from the microscopic theory by integrating out the microscopic degrees of freedom from the microscopic action of the system.

We consider a system described by the tunneling Hamiltonian of the form
\begin{equation}
H = H_1 + H_2 + H_T ,
\end{equation}
where $H_1$ and $H_2$ are the Hamiltonians describing two independent systems and $H_T$ is the tunneling Hamiltonian that parametrizes the interaction between them via tunneling matrix elements $T_{n}$, of a particle on state $n$ tunneling between the systems. In this case the effective action is given by \cite{snyman2008keldysh}
\begin{equation}
\mathcal{S}_{\rm{eff}} = \frac{1}{2} \sum_{n} \int \frac{d\epsilon}{2 \pi} \rm{Tr} \rm{ln} \left[1+ T_n \frac{\{\check{G}_1, \check{G}_2\} - 2}{4} \right] ,
\label{connectoraction}
\end{equation}
where $\check{G}_{L/R}$ are the quasiclassical Keldysh Green's functions of the two uncoupled systems.

For a mesoscopic island connected to two electron reservoirs (the left and the right lead), we can use this formalism to write the partition function for the heat current and the charge current to the island as \cite{laaksothesis} 
\begin{equation}
\begin{aligned}
& \mathcal{Z} =  \int \mathcal{D} \chi \int \mathcal{D} \xi \int \mathcal{D} Q \int \mathcal{D}E \\
&  \times \exp \left\{-\int dt\left[\xi \dot{E} + \chi \dot{Q} - S_{\rm{eff},L}(\xi, \chi) - S_{\rm{eff},R}(\xi, \chi)\right]\right\} .
\label{partition}
\end{aligned}
\end{equation}
Here the first two terms in the exponent include the energy, $E$, and charge, $Q$, on the island along with their corresponding counting fields, $\xi$ and $\chi$, and are introduced to ensure the charge and energy conservation on the island. The two connector actions describe the left and the right contacts. In interacting systems, the partition function often contains also functional integrals over some fields that are the result of Hubbard-Stratonovich decoupling of the interactions. Below we assume that they are evaluated using the saddle point approximation.

For tunneling contacts with $T_n \ll 1$ we can expand the action in tunneling matrix elements. To first order, the total connector action becomes
\begin{equation}
\mathcal{S}_{\rm{eff}} \approx \frac{1}{8} \sum_\alpha g_\alpha \int \frac{d\epsilon}{2\pi} \rm{Tr} \left[\{\check{G}_1, \check{G}_2\} - 2\right],
\end{equation}
where $g_\alpha=\sum_n T_n$. For simplicity, we concentrate only on this term below and assume symmetric contacts, $g_L=g_R=2\pi G_T/e^2$ with the tunneling conductance $G_T$. 

We proceed by substituting Green's functions in the action. As a result, the action becomes in the first order in the tunneling matrix elements \cite{laakso2012theory}
\begin{equation}
\begin{aligned}
\mathcal{S}_{\rm{eff}} = & \int d\epsilon \Big[ (e^{\chi(t) + \xi(t) \epsilon} - 1) \gamma^+(\epsilon, t) \\
& + (e^{-\chi(t) - \xi(t) \epsilon} - 1) \gamma^-(\epsilon, t) \Big].
\end{aligned}
\label{action11}
\end{equation}
Here $\gamma^\pm$ are the probabilities per unit time and energy of particles on energy state $\epsilon$ tunneling at time $t$ into ($+$) or out of ($-$) the island. The tunneling rates are given by $\Gamma^\pm = \int d\epsilon \gamma^\pm$. For example, in the case of a superconducting island coupled to normal metal leads, $\gamma^\pm$ are given by
\begin{equation}
\gamma^+(\epsilon) = \frac{G_T}{e^2} N_S(\epsilon) (f_L(\epsilon) + f_R(\epsilon)) (1-f_I(\epsilon)) ,
\label{gammaplus}
\end{equation}
\begin{equation}
\gamma^-(\epsilon) = \frac{G_T}{e^2} N_S(\epsilon) (2-f_L(\epsilon) - f_R(\epsilon)) f_I(\epsilon) .
\label{gammaminus}
\end{equation}
Here $N_S$ is the BCS density of states and $f_{I/L/R}$ are the distribution functions on the island and the left and the right leads, respectively

Below we consider a system with some general rates $\gamma(\epsilon)$. The considerations could also be generalized to higher-order processes in $T_n$ \cite{laakso2012theory}.

\section{Langevin equation from the action}

\subsection{Discrete Langevin equation}

We now show how Eq. \eqref{partition} leads to the Langevin equation for the energy of the island. For this, we neglect the charge counting field $\chi$.

Expanding the integrand in the partition function in the powers of $\gamma^\pm$ gives
\begin{equation}
\begin{aligned}
\mathcal{Z} = & \int \mathcal{D} E \mathcal{D}\xi \sum_{n,m=0}^\infty \Bigg[\prod_{k,j=1}^{n,m} \int dt_k d\epsilon_k dt'_j d\epsilon'_j e^{-\int dt d\epsilon \gamma^+ + \gamma^-} \\
& \times \frac{\gamma^+(t_k,\epsilon_k) \gamma^-(t'_j,\epsilon'_j)}{m!n!} \Bigg] \exp\Bigg[-\int dt \xi(t) \\
& \times \Bigg( \dot{E} - \sum_k \epsilon_k \delta(t-t_k) + \sum_j \epsilon'_j \delta(t-t'_j)\Bigg)\Bigg] .
\end{aligned}
\end{equation}
The exponential term with the sum over the delta functions follows from the relation
\begin{equation}
\xi(t_n)= \int dt \xi(t) \delta(t-t_n) .
\end{equation}
This form allows us to carry out the functional integration over the counting field $\xi$. This results in a delta function that indicates that only the paths that satisfy the equation
\begin{equation}
\dot{E} = \sum_k \epsilon_k \delta(t-t_k) - \sum_j \epsilon'_j \delta(t-t'_j)
\label{langevin1}
\end{equation}
contribute to the partition function. The energy of the island is thus changed in discrete steps by particles tunneling at times $t_i^{(\prime)}$ with energies $\epsilon_i^{(\prime)}$. The distribution of the times and energies of these tunneling events are given by the weighting factor in the partition function. In this case they are Poisson distributed.

\subsection{Continuum Langevin equation with uncorrelated tunneling events}

From Eq. \eqref{langevin1} describing discrete tunneling events we can move to a continuum description as follows. Let us consider the total energy transferred to the island during some long time interval of length $T \gg 1/\Gamma^\pm$. Let us further divide this time to shorter time intervals $\Delta t \ll T$. The rate of change of the energy during this longer time interval $T$ is given by
\begin{equation}
\begin{aligned}
\dot{E} = & \frac{1}{T} \sum_n \epsilon_n \sum_{m=1}^{T/\Delta t} \left(n^+(\epsilon_n, t_m) - n^-(\epsilon_n, t_m)\right) .
\end{aligned}
\end{equation}
 Here $n^\pm(\epsilon_n,t_m)$ is a random variable that gives the number of $+$ or $-$ tunneling events during the time interval $[t_m, t_m+\Delta t]$ and energy interval $[\epsilon_n, \epsilon_n + \Delta\epsilon]$. As shown above, the number of events in this two-dimensional interval is Poisson distributed with mean and variance of $\Delta t \Delta \epsilon \gamma^\pm(\epsilon, t)$. Now we define a new random variable as
\begin{equation}
N(\epsilon_n) = \sum_m \left(n^+(\epsilon_n, t_m) - n^-(\epsilon_n, t_m) \right) .
\end{equation}
If we can assume that the tunneling events at different times are independent of each other, the central limit theorem tells us that the random variable $N(\epsilon)$ is normally distributed as
\begin{equation}
\begin{aligned}
N(\epsilon) \sim \mathcal{N}\Bigg(&\Delta \epsilon \int dt (\gamma^+(\epsilon,t)-\gamma^-(\epsilon,t)),\\
& \Delta \epsilon \int dt (\gamma^+(\epsilon,t)+\gamma^-(\epsilon,t)) \Bigg) ,
\end{aligned}
\end{equation}
where the first argument is the mean and the second argument is the variance. We have also taken the limit $\Delta t\rightarrow 0$ and converted the sum over times to an integral. The total energy transferred during the time interval $[t,t+T]$ becomes a linear combination of normally distributed random variables with coefficients $\epsilon_i$. A linear combination of normal random variables is also a normal random variable and thus the total energy transferred during the time interval $[t,t+T]$ is given by
\begin{equation}
\begin{aligned}
\Delta E(t) \sim \mathcal{N} \left(T \dot{H},T S_{\dot{H}} \right) ,
\end{aligned}
\end{equation}
where we have also approximated the rates in the time integral by constants in time. The mean and the variance of the distribution are given in terms of the energy current to the island, $\dot{H} = \int d\epsilon \epsilon \left(\gamma^+(\epsilon,t) - \gamma^-(\epsilon,t) \right)$, and its noise, $S_{\dot{H}} = \int d\epsilon \epsilon^2 \left(\gamma^+(\epsilon,t) + \gamma^-(\epsilon,t) \right)$. Dividing by $T$ and separating the random variable to stochastic and deterministic parts, we get the Langevin equation for the energy of the island
\begin{equation}
\dot{E} = \dot{H}(E(t)) + \xi(t) ,
\label{langevin2}
\end{equation}
where $\xi(t)$ is a normal random variable with zero mean and variance $S_{\dot{H}}$.

In order to be able to derive Eq. \eqref{langevin2}, we had to assume that the numbers of tunneling events on the small time intervals are uncorrelated. Since the tunneling probabilities depend on the state of the island and the tunneling events change the state, this assumption is justified only in the limit where the change of the state due to one tunneling event is negligible, i.e., when there are no effects due to the charging energy, (i.e. $E_C \ll k_B T$), and the energy change of the island due to one tunneling event is small compared to the total energy of the island. For a quantum dot with level spacing $\delta \sim k_B T$, the assumption is also broken because the tunneling of one particle to the island heavily affects subsequent tunneling events due to the Pauli exclusion princple. We note that Eq. \eqref{langevin2} can also be derived directly from the action by expanding it to the second order in $\xi$ and then carrying out functional integration over $\xi$ \cite{laakso2012theory}. Then the assumption of uncorrelated tunneling events means that we can disregard terms of higher order than $\xi^2$.

\section{Application to a NISIN junction}

For a superconducting island there are several processes that in principle contribute to the state of the island. The above discussion can be used to take into account the process where single quasiparticles tunnel into or out of the island. Processes of a higher order in the tunneling matrix elements, such as Andreev reflection and co-tunneling, can be neglected if we consider only contacts with low transparency. Other processes include quasiparticle scattering from phonons, which can be neglected choosing materials with small electron-phonon coupling constant and limiting the consideration to low temperatures. One more possible process is the quasiparticle recombination process which also is typically driven by electron-phonon coupling. The rate for this process is analyzed in Ref. \cite{kaplan1976quasiparticle}. For quasiparticles at the gap edge, $\epsilon \sim \Delta$, the tunneling rate divided by the recombination rate is given by 
\begin{equation}
\frac{\Gamma_T}{\Gamma_r} \sim 400\,000 \times \frac{\left(\frac{\tau_0}{440 \times 10^{-9} \,\rm{s}}\right) \left(\frac{\Delta}{k_B \times 2 \,\rm{K}}\right)}{\left(\frac{R}{1 \,\rm{k\Omega}}\right)} ,
\end{equation}
where $\tau_0$ is the characteristic recombination time in aluminium and $R$ is the tunnel resistance. The recombination rate can be neglected ($\Gamma_T \gg \Gamma_r$ for $R \ll 0.4\,\rm{G\Omega}$.

\subsection{Gaussian limit}

Let us consider the case when the energy relaxation time for the quasiparticles on the island is faster than the injection rate. In this case the quasiparticles on the island can be described by the Fermi-Dirac distribution with some fluctuating temperature $T$. From the BCS theory, the energy of a superconducting island at equilibrium for $k_B T \ll \Delta$ is given by
\begin{equation}
E = \frac{\sqrt{2 \pi \Delta^3 k_B T}}{\delta} e^{-\frac{\Delta}{k_B T}} ,
\label{thermalenergy}
\end{equation}
where $\delta$ is the average level spacing of the island. In all considerations below, we assume $\delta \ll k_B T, \Delta$. On the other hand, the energy carried by a single tunneling quasiparticle (for bias voltage $\vert e V\vert < \Delta$) is of the order of $\Delta$. In the light of the discussion in the previous section, the criterion for the independence of tunneling events is thus given by $\Delta \ll E$. This can also be written as $N_{qp} \gg 1$, where $N_{qp}$ is the mean quasiparticle number on the island. We consider this limit first. Furthermore, by assuming that there is no charge imbalance on the island, which is valid for a left-right symmetric setup, we can choose the chemical potential of the island to zero so that the energy transferred to the island coincides with the heat transferred to the island.

It is most convenient to consider the fluctuations of thermodynamic quantities in terms of the probability distribution for the island being in some given state. The Langevin equation with Gaussian noise can be related to the probability distribution $P(E)$ for having the island at some energy or, equivalently, the probability distribution $P(T)$ for the temperature of the island. This can be solved from the corresponding Fokker-Planck equation (for derivation, see \cite{laakso2010giant})
\begin{equation}
\dot{H} P(T) = \frac{1}{2} \partial_T \left(C^{-1} S_{\dot{H}} P(T) \right) ,
\end{equation}
where $C$ is the heat capacity of the island. Solving this gives
\begin{equation}
P(T) \propto \exp\left[\int dT \left( \frac{2 C(T) \dot{H}(T) - \partial_T S_{\dot{H}}(T)}{S_{\dot{H}}(T)} + \frac{\partial_T C(T)}{C(T)}\right)\right] .
\label{fokkerplanck}
\end{equation}
By substituting into the heat current and its noise the tunneling probabilities from Eqs. \eqref{gammaplus} and \eqref{gammaminus} and analyzing the above expression, we see that in the limit where the tunneling events are uncorrelated, the last two terms in Eq. \eqref{fokkerplanck} can be neglected. Furthermore, approximating the heat capacity as a constant and evaluating the heat current and its noise in the first order in $k_B T/\Delta$, we get
\begin{equation}
P(T) \propto \exp\left[- \frac{C(T_N)}{2 k_B T_N^2} \left(T_S-T_{\rm{eff}} \right)^2\right] .
\end{equation}
This is the Gaussian distribution, as expected for equilibrium fluctuations \cite{landau1996statistical}. What is more interesting is that the distribution keeps the Gaussian form (within the low temperature approximation) even when the system is biased with voltage $\vert e V \vert \ll \Delta$. The bias can be taken into account in 
\begin{equation}
T_{\rm{eff}} = \frac{2\Delta}{W\left(\frac{2 \Delta}{k_B T_N} \exp\left(\frac{2 \Delta - eV}{ k_B T_N}\right)\right) } ,
\end{equation}
where $W(x)$ is the Lambert W-function.

\subsection{Non-Gaussian limit}

Next we consider the superconducting island in the limit where the tunneling events are correlated, i.e., $N_{qp} \sim 1$ and $\Delta \approx E$. Because the tunneling rates are larger than other rates that change the quasiparticle number of the island, the quasiparticle number is a well-defined quantity between the tunneling events and we can consider a Master equation approach to calculate the statistics of the quasiparticle number. For now, we stick to the assumption that the quasiparticles can relax to the equilibrium distribution immediately after tunneling. We return to the validity of this assumption below. As an approximation, we assume that each tunneling event carries energy $\Delta$. This approximation can be validated by evaluating the effective action \eqref{action11} in the lowest order in the temperature in a similar way as in Refs.~\cite{anghel:197,laakso2012theory}. As a result, the two terms describing the tunneling into and out of the island become proportional to $\sim \exp(\pm\xi \Delta)$. Differentiating this with respect to the counting field, we find that for example the heat current is given by $\Delta$ times the particle current into the island, i.e. each tunneling particle increases or decreases the energy of the island by $\Delta$.

The effect of the tunneling on the state of the island is to change the thermal energy on the island which in turn has a one-to-one correspondence with the temperature of the island. The temperature after tunneling can be calculated by inverting Eq. \eqref{thermalenergy} and substituting for the thermal energy $E = N_{qp} \Delta$. Using the formalism of the previous section, the Langevin equation for the energy of the island becomes
\begin{equation}
\dot{E} = \sum_n \Delta \delta(t-t_n) - \sum_m \Delta \delta(t-t_m) ,
\label{langevinNISIN}
\end{equation}
where $t_n$ and $t_m$ are Poisson distributed with rates $\Gamma^+$ and $\Gamma^-$, respectively. Written explicitly, the rates read to the lowest order in temperature and in the limit $\vert\mu_\alpha\vert < \Delta$
\begin{equation}
\Gamma^+_\alpha = \frac{G_T}{e^2}\sqrt{\frac{2 \Delta k_B T_N}{\pi}} e^{-\frac{\Delta - \mu_\alpha}{k_B T_N}}
\label{g1}
\end{equation}
and
\begin{equation}
\Gamma^-_\alpha = \frac{G_T}{e^2}\sqrt{\frac{2\Delta k_B T_S}{\pi}} e^{-\frac{\Delta}{k_B T_S}} .
\label{g2}
\end{equation}
Here $T_N$ is the temperature of the leads, $T_S$ is the fluctuating temperature of the island corresponding to the energy of the island $E = N \Delta$ and the chemical potential is given by $\mu_L=e V / 2$ and $\mu_R = -e V / 2$ for the left and the right leads, respectively. Due to our approximation of each particle carrying energy $\Delta$, this can also be interpreted as a Langevin equation for the particle number on the island. The corresponding master equation with rates \eqref{g1} and \eqref{g2} gives the probability distribution for the quasiparticle number on the island, which is shown in Fig. \ref{fig:langevinNISIN}. 

For a low average $N_{qp}$, the independent tunneling approximation breaks down. This can be seen in the fact that even in equilibrium the average $N_{qp}$ obtained from Eq. \eqref{fokkerplanck} does not equal
\begin{equation}
N_{qp}(T \ll \Delta) = \frac{2}{\delta} \sqrt{2 \pi \Delta k_B T} e^{-\frac{\Delta}{ k_BT}} 
\label{nqpbcs}
\end{equation}
obtained from the BCS formula. Rather, one has to use the distribution obtained from Eq.~\eqref{langevinNISIN}.


Above we assume the island to be in quasiequilibrium, i.e. that the distribution function of the island can be described by an effective temperature $T_I$. Whether this is a good assumption is debatable. In the lowest order in temperature, we can, however, show that this assumption plays no role, because it gives the same rates as we would get by just assuming that all quasiparticles are distributed close to the gap edge. We now present an argument to derive the same rates without assuming a thermal distribution on the island.

For tunneling from lead $\alpha$ to the island with only a few excited quasiparticles, the rate is only determined by the filling factor in the lead. This is because for a small number of excitations on the island, the number of available states close to the gap edge is much larger than the number of excitations due to the BCS divergence in the density of states and a small level spacing. Thus the tunneling rate from lead $\alpha$ to the island equals
\begin{equation}
\Gamma^+_\alpha  = \frac{G_T}{e^2} \int_0^\infty d\epsilon N_S(\epsilon) f^\alpha(\epsilon-\mu_\alpha) ,
\end{equation}
which gives Eq. \eqref{g1} in the low-temperature limit. For tunneling from the island to lead $\alpha$, the rate is determined by the number of excitations on the island. From the Fermi golden rule, assuming that all quasiparticles are located at energy $\Delta$ (and $\vert\mu_\alpha\vert < \Delta$), we get
\begin{equation}
\Gamma^-_\alpha = \frac{G_T}{4 e^2} N \delta .
\end{equation}
This is equivalent to \eqref{g2}, provided that we can substitute for the particle number an effective temperature defined by the BCS formula \eqref{nqpbcs}.

In the discussion above, we did not consider the effect of possible fluctuations of the superconducting gap $\Delta$. From the standard BCS gap equation one can show that the corrections to the zero-temperature gap with $N$ particles on energy state $\Delta$ are of the order of $\sim N \delta$, which means that they can be neglected. On the other hand one could imagine that the modifications to the partition function \eqref{partition} change the gap equation somehow. However, these modifications are proportional to the tunneling rates, which are small compared to the terms appearing in the BCS energy term. The latter term alone gives the BCS gap equation.

\section{Conclusion}

We have shown how the Langevin (or equivalently, the Master or Fokker-Planck) equation can be derived from the effective action describing particle transport in a setup consisting of a mesoscopic island and two leads. We show how the Langevin-like equation describing discrete tunneling events leads to the traditional Langevin equation with continuous transport rate in the limit of uncorrelated tunneling events. We have also applied the formalism to calculate the temperature and quasiparticle excitation statistics on a superconducting island coupled to two normal metal leads.

\section*{Acknowledgment}
This work was supported by the Academy of Finland Center of Excellence program, European Research Council (Grant No. 240362-Heattronics), EU-FP 7 MICROKELVIN (Grant No. 228464) and INFERNOS (Grant No. 308850) programs.



\bibliographystyle{IEEEtran}

%



\end{document}